\documentclass[aps,amsfonts,12pt]{article}
\usepackage{feynmp}
\usepackage{appendix}
\usepackage{graphicx}
\usepackage{ dsfont }
\usepackage{amssymb}
\usepackage{ tipa }
\usepackage{amsmath}
\usepackage{amsbsy}
\usepackage{a4wide}
\usepackage{color}
\usepackage{slashed}
\usepackage{amsmath}
\usepackage{amsbsy}
\usepackage{bm}
\usepackage{color}
\usepackage{epsfig}
\usepackage{float}
\usepackage{ulem}
\newcommand{\ee}{\end{equation}}
\newcommand{\bb}{\begin{equation}}
\newcommand{\eqb}{\begin{eqnarray}}
\newcommand{\eqf}{\end{eqnarray}}

\usepackage{upgreek}
\usepackage{ stmaryrd }
\definecolor{gre}{rgb}{0,0.4,0.3}

\usepackage{color}

\begin{document}
\title{Mixing of photons with light pseudoscalars in time-dependent magnetic fields}
\author{Paola Arias$^{a,}$\footnote{{
e-mail}: paola.arias.r@usach.cl } \,\,, Ariel Arza$^{a,}$\footnote{{
e-mail}:ariel.arza@usach.cl }\,\,, and Jorge Gamboa$^{a,}$\footnote{{
e-mail}: jgamboa55@gmail.com}\,\,
\\[2ex]
\small{ $^a$Departmento de F\1sica, Universidad de Santiago de Chile, Casilla 307, Santiago, Chile}}

\date{}

\maketitle

\begin{abstract}
The effects of an external time-dependent magnetic field in the conversion probability of photon-to-axion-like particles are studied. Our findings show that  for a certain time regime, the amplitude of the produced axion-like field can be enlarged with respect to the static case, thus, enhancing the probability of conversion.

\end{abstract}
\maketitle

\section{Introduction}

Particles beyond the Standard Model (SM) are frequently invoked to solve some tensions or fine tuning issues of the model. The most iconic example being the absence of a dark matter candidate, where usually proposals like Weakly Interacting Massive Particles (WIMPs) and Weakly Interacting Slim Particles (WISPs) take the lead, but some other exotic options have also been put forward \cite{Bertone:2004pz}. Another good example where new particle content has been invoked is to  explain  the absence of CP violation in strong interactions {\it a.k.a.} the  strong CP problem \cite{peccei}. The most accepted solution to this problem is the existence of a new U(1) symmetry, the Peccei-Quinn symmetry, which is spontaneously broken at very high energies $f_a$. The new Nambu-Goldstone emerging from the spontaneous breaking is known as the axion particle. They acquire a very small mass due to the explicit breaking of the Peccei-Quinn symmetry when the QCD instanton effects turn on. Thus, there is a close relationship between $f_a$ and the mass of the axion \cite{Agashe:2014kda}, namely  $m_\phi f_\phi \sim m_\pi f_\pi$, where $m_\phi$ is the pion mass and $f_\pi$ its decay constant. Axions also develop a coupling to two photons of the form
\bb
\mathcal L_{a\gamma\gamma} = g_{a\gamma\gamma} a F_{\mu\nu}\tilde F^{\mu\nu}, \label{alpcoupling}
\ee
via pion mixing. Here $g_{a\gamma\gamma}$ is the coupling strength of axions to photons, $a$ the axion field and $F_{\mu\nu} (\tilde F_{\mu\nu})$ the electromagnetic (dual) tensor. 

After the proposal of the axion mechanism to naturally explain the absence of CP violation in the strong sector, many efforts have been made to motivate the existence of new scalar bosons with similar characteristic to the axion from a bottom-up approach \cite{witten1, witten2}. For instance in \cite{witten1} has been shown that every string theory has at least one zero-mode or axion-like particle  (ALP so forth) in the fourth-dimensional reduction. These new (pseudo) scalar bosons ($\phi$) can also develop couplings such as the one in eq.~(\ref{alpcoupling}), but they do not feature the {\it a priori} relationship between the coupling and the mass ($m_{\phi})$. 

Thus, new pseudo-scalar bosons can be searched for exploiting the coupling to two photons, specially via the Primakoff effect: a photon beam is sent transversally to a region with a strong electric or magnetic field (sea of virtual photons) so the conversion (oscillation) of photons into ALPs can take place. For detection purposes, it is needed to have a second identical region, impermeable to photons, so the re-conversion can take place and a photon is detected \cite{Sikivie:1983ip}.
For such a configuration, the probability of conversion of a photon into an ALP ($\phi$) and then back into a photon in the presence of a constant magnetic field strength $B_0$, is given by \cite{detailed_sikivie}
\bb
P_{\gamma\rightarrow \phi \rightarrow\gamma}= \frac{1}{16} \frac{\omega^2}{k_{\phi}^2} \left(g B_0 L\right)^4 \frac{\mathcal F_r}{\pi} \frac{\mathcal F_p}{\pi} \left| \frac{2}{qL}\sin\left(\frac{qL}{2} \right)\right|^4
\label{probabilityALP}\ee
Where $\omega$ is the frequency of the incoming photon, $L$ is the length of the magnetic region, $k_\phi=\sqrt{\omega^2 -m_{\phi}^2}$ and $q=\omega-k_\phi\approx m_\phi^2/(2\omega)$ is the momentum transfer. We have included the so-called finesse of the cavity ($\mathcal F$) of the production and regeneration side $\mathcal F_p, \mathcal F_r$, respectively for completeness. The above formula assumes the spatial extent of the photon beam transverse to its direction is much bigger than the wavelength, thus photons effectively propagate in one spatial dimension (say $\hat x$) and the magnetic field it is homogeneous and directed  along the $\hat z$ direction, thus the created ALP propagates in the same direction as the photon. As can be seen from eq.~(\ref{probabilityALP}), the  use of  strong magnetic fields and large optical paths enhance the probability of conversion. Also, spatially inhomogeneous magnetic fields have been considered as a way of enhance the oscillation \cite{adler2, Guendelman}. The enlargement of the optical path of the incoming beam of photons can be achieved by placing mirrors  (Fabry-Perot cavities \cite{hoogeveen, sikivie-bibber})  in both production and regeneration regions or use shielded microwave cavities \cite{hoogeveen-RF, jaeckel}, thus imposing boundary conditions for the incoming electric field.  Some experimental searches using the conversion technique discussed above are \cite{experiments}.

In this paper we are interested in studying the effects of time-dependent magnetic fields on the conversion probability of new scalars that share the coupling to two photons given by eq.~(\ref{alpcoupling}) via Primakoff effect. So far, up to our knowledge, time-dependent magnetic fields have not been studied extensively in the literature \cite{Adler:2006zs},  however, interesting properties, such as new types of resonances, could appear allowing to amplify the conversion probability. 

Our study is intended to stay in a simplified one dimensional scenario, as the case described above. Nonetheless our findings could drive the attention to, in the future, perform a detailed analysis of the use of time-dependent magnetic fields in laboratory-based searches.

The manuscript is organized as follows: In section \ref{setup} we give explain the setup in the production region, where the conversion photon-ALP (and viceversa) takes place and we set the equations of motion . We solve them using proper initial and boundary conditions and find the probability to regenerate a photon in the second cavity. Finally in section \ref{conclusions} we comment on the validity of our results and possible future applications.

\section{Photon-ALP conversion in a time-dependent magnetic field}\label{setup}

The action to be considered is given by

 \eqb
\mathcal{S}=\int d^4x\left[-\frac{1}{4}F_{\mu\nu}F^{\mu\nu}+\frac{1}{4}g\phi F_{\mu\nu}\tilde F^{\mu\nu}+\frac{1}{2}(\partial_\mu \phi)^2-V(\phi)\right], \label{0}
\eqf
{{ where $F_{\mu\nu}$ is the electromagnetic field strength, $\tilde F^{\mu\nu}$ its dual,  $\phi$ is the ALP field, $g$ the coupling strength to photons, {{and $V(\phi)$ the ALP potential is usually parametrized as
\bb
V(\phi)=m_\phi^2f_\phi^2\left[1-\cos\left(\frac{\phi}{f_\phi}\right)\right]=m_\phi^2f_\phi^2\left[\frac{1}{2!}\left(\frac{\phi}{f_\phi}\right)^2-\frac{1}{4!}\left(\frac{\phi}{f_\phi}\right)^4+...\right]. \label{potential}
\ee 
 Here $f_\phi$ is the decay constant of the {{scalar}} field and is related with the coupling to photons  by $g={\alpha}/({2\pi f_\phi})$, where $\alpha$ is the fine structure constant.}}

In the present   setup  we consider the situation where photons are enclosed  in a region (production)  with two infinite non-ideal parallel plates, positioned at $x=0$ and $x=L$, respectively. They will impose boundary conditions for the vector potential given by $A(0,t)=A(L,t)=0$,  (it could be, for instance, a Fabry-Perot cavity). We assume that once a standing wave is formed inside the cavity, suddenly, in $t=0$, an external time-dependent magnetic field, $B_\text{ext}\hat z$, is turned on with a periodic dependence on time, such  that $B_{\rm ext}=B_0 \cos( \lambda t)$  \footnote{We will not consider the effects of the induced electric field that will produce the time-dependent magnetic field, assuming that their effect can be attenuated with respect to the magnetic field.}.  There is an identical second region (regeneration), between $x=L$ and $x=2L$ where the reconversion can take place.

In the Coulomb gauge we have the following equations of motion
\eqb
\left(\partial_t^2-\partial_x^2\right)A(x,t)& = & -gB_\text{ext}(t)\partial_t\phi(x,t), \label{1} 
\\
\left(\partial_t^2-\partial_x^2+m_\phi^2\right)\phi(x,t) &=& gB_\text{ext}(t)\partial_tA(x,t), \label{2}
\eqf
where $A$ corresponds to $\hat z$ component of $A^\mu$. 

Our strategy to solve the coupled eqs.~(\ref{1}) and (\ref{2}) is to consider that firstly a standing wave is formed inside the production region
\bb
A^{(0)}_n(x,t)=a\sin\omega_nt\sin\omega_nx, \label{solA0}
\ee
where $\omega_n=n\pi/L$ is the resonant frequency mode ($n \in \mathds{Z}$) and $a$ is the amplitude of electromagnetic standing wave, which is  related with the power stored (or emitted) by the cavity, ${\cal P}$, as  
\bb
{\cal P}=\frac{|a|^2}{4} \frac{S \omega_n^3 L}Q, \label{a}
\ee
where $Q$ is the quality factor of the cavity and $S$ is the cross sectional area of the cavity mode. The quality factor is related to the finesse of the cavity (in vacuum) as $\mathcal F=\pi Q/(L\omega)$. We assume the frequency of the excited mode in both production and regeneration regions are the same, and tuned such  that the frequency of the incoming photon, $\omega$, matches the frequency of the resonant mode $\omega_n$ inside the cavities.

Secondly,  at first order in $g$, this standing wave will act as a  source for the ALP field,  $\phi^{1}_n(x,t)$, in the production region. The scalar field propagates undamped to the second regeneration region (identical to the production region) and analogously, acts as a source to regenerate the photon field  $A^r(x,t)$. Finally the probability to reconvert a photon in the second cavity can be computed using the power  stored in each production ($\mathcal P_p)$ and regeneration ($\mathcal P_r$) cavities
\bb
P_{\gamma\rightarrow \phi\rightarrow \gamma}= \frac{\mathcal P_r}{\mathcal P_p}.
\label{probability}
\ee
So we put eq.~(\ref{solA0}) back into eq.~(\ref{2}).

\bb
\left(\partial_t^2-\partial_x^2+m_\phi^2\right)\phi^{(1)}_n(x,t)=f_n(t)\Theta(x)\Theta(L-x)\sin(\omega_nx), \label{eqphi1}
\ee
where
\bb
f_n(t)=\frac{gB_0\omega_na}{2}\left(\cos(\Omega_nt)+\cos(\bar\Omega_nt) \right), \label{f}
\ee
and $\Omega_n=\omega_n+\lambda$, $\bar\Omega_n=\omega_n-\lambda$.  The function $\Theta(y)$ is the Heaviside step function, defined as $\Theta(y)=1$ for $y>0$ and $\Theta(y)=0$ for $y<0$.
Let us note that we have considered only the first term of the ALP potential given in eq.~$(\ref{potential})$, thus retaining only the linear contribution. We shall discuss later about the effects of considering non-linear terms. 

In order to  solve for $\phi^{(1)}_n(x,t) $ we  perform a Laplace transform to eq.~(\ref{eqphi1})   with initial conditions $\phi^{(1)}_n(x,0)=0$ and $\dot\phi^{(1)}_n(x,0)=0$. 
We define as
\bb
\Phi^{(1)}_n(x,s)=\int_0^\infty dt \ e^{-st}\phi^{(1)}_n(x,t), \label{LTphi}
\ee
and
\bb
F_n(s)=\int_0^\infty dt \ e^{-st}f_n(t), \label{LTf}
\ee
the Laplace transforms of $\phi^{(1)}_n(x,t)$ and $f_n(t)$, respectively, and now eq.~(\ref{eqphi1}) takes the form 
\bb
\left[\partial_x^2-\eta^2(s)\right]\Phi^{(1)}_n(x,s)=-F_n(s)\Theta(x)\Theta(L-x)\sin(\omega_nx), \label{eqPhi1}
\ee
where we have defined $\eta(s)=\sqrt{s^2+m_\phi^2}\equiv \eta_s$. 

This equation can be solved by the Green method
\bb
\Phi^{(1)}_n(x,s)=-F_n(s)\int_{-\infty}^\infty dx' G(x,x';s)\Theta(x')\Theta(L-x')\sin(\omega_nx'), \label{solPhiGreen}
\ee
where $G(x,x';s)$ is the Green function given by
\bb
G(x,x';s)=-\frac{1}{2\eta_s}e^{-\eta_s |x-x'|}. \label{green}
\ee
The complete solution for $\Phi^{(1)}_n(x,s)$ becomes 
\bb
\Phi^{(1)}_n(x,s) =
\begin{cases}
\frac{\omega_n}{2\eta_s}\frac{F(s)}{s^2+k_n^2}\left({e^{-\eta_s x}-(-1)^ne^{-\eta_s (x-L)}}\right);   & x\geq L,  \\
 \frac{\omega_n}{2\eta_s}\frac{F(s)}{s^2+k_n^2}\left({e^{\eta_s x}-(-1)^ne^{\eta_s (x-L)}}\right);   & x\leq 0, \\
\frac{\omega_n}{2\eta_s}\frac{F(s)}{s^2+k_n^2}\left[\left({e^{-\eta_s x}-(-1)^ne^{\eta_s (x-L)}}\right)+\frac{2}{\omega_n}\sin(\omega_nx)\right]; & 0<x< L,
\end{cases}
\ee
where $\kappa_n=\sqrt{\omega_n^2+m_\phi^2}$.

To obtain the solution for the ALP field in the time space, we apply the inverse Laplace transform  ${\cal L}^{-1}[\Phi^{(1)}_n(x,s)]$, 
\bb
\phi^{(1)}_n(x,t) =
\begin{cases}
\frac{gB_0\omega_n^2a}{4}\left[{\mathcal J}_n(x,t)-(-1)^n{\mathcal J}_n(x-L,t)\right]; & x\geq L, 
\\
\\
\frac{gB_0\omega_n^2a}{4}\left[{\mathcal J}_n(-x,t)-(-1)^n{\mathcal J}_n(L-x,t)\right]; & x\leq 0,
\\
\\
\frac{gB_0\omega_n^2a}{4}\left[{\mathcal J}_n(x,t)-(-1)^n{\mathcal J}_n(L-x,t)+\frac{2}{\omega_n}\xi_n(t)\sin\omega_nx\right]; & 0<x<L,
\end{cases}
\label{PhiTransf}
\ee
where the function ${\mathcal J}_n(x,t)$ is defined as
\bb
{\mathcal J}_n(x,t)=\int_0^tdt'\xi_n(t-t')\Theta(t'-x)J_0\left(m_\phi\sqrt{t'^2-x^2}\right), \label{J}
\ee
with $J_0(x)$ the Bessel function of first kind, and
\bb
\xi_n(t)=\frac{\cos(\Omega_nt)-\cos (k_nt)}{k_n^2-\Omega_n^2}+\frac{\cos(\bar\Omega_nt)-\cos (k_nt)}{k_n^2-\bar\Omega_n^2}. \label{xi}
\ee

From the above equation, there is a resonance for the ALP field if the parameter $\kappa_n$ takes the values $\Omega_n$ or $\bar\Omega_n$. Let us choose $\kappa_n=\Omega_n$ and indeed we find
\bb
\xi_n(t)=\frac{t\sin(\Omega_nt)}{2\Omega_n}+\frac{\sin\left(\lambda t\right)\sin(\omega_nt)}{2\lambda\omega_n}. \label{xi1}
\ee
Note that the first term on the right-hand side has its amplitude enhanced by $t$. Therefore, in order to have such enhancement the frequency of the external magnetic field should be tuned such as to match the momentum transfer (assuming $\omega_n \gg m_\phi)$
\bb
\lambda=\sqrt{\omega_n^2+m_\phi^2}-\omega_n\approx \frac{m_\phi^2}{2\omega_n}. \label{lambda}
\ee 
For operating times of the magnetic field that satisfy  $t\gg1/\lambda$ (which seems a plausible assumption) the first term in eq.~(\ref{xi1}),  will dominate. Retaining only this term, eq.~(\ref{J}) becomes
\bb
{\mathcal J}_n(x,t)=\frac{1}{2\Omega_n}\int_0^tdt'(t-t')\sin(\Omega_n(t-t'))\Theta(t'-x)J_0\left(m_\phi\sqrt{t'^2-x^2}\right). \label{J1}
\ee

To get an analytical expression for $\phi^{(1)}_n (x,t)$, we can integrate the above equation for small masses of the ALP field, namely  $m_\phi\ll\Omega_n$. In this case, the Bessel function $J_0\left(m_\phi\sqrt{t'^2-x^2}\right)$ has a slower oscillation frequency than  the term $(t-t') \sin(\Omega_n(t-t'))$ in the integration range $[x,t]$. Therefore, we integrate as follows
\eqb
{\mathcal J}_n(x,t)& \approx & \Theta(t-x)\left[J_0\left(m_\phi\sqrt{t'^2-x^2}\right) \frac{1}{2\Omega_n}\int dt'(t-t')\sin(\Omega_n(t-t')) \right]_{t'=x}^{t'=t}, \nonumber  \\ 
& \approx &  -\frac{1}{2\Omega_n^2}\Theta(t-x)(t-x)\cos(\Omega_n(t-x)). \label{J2}
\eqf
Where we have retained only the resonant term.

Now the ALP field can leave the first region and propagate into the second one, where $x\geq L$. By replacing eq.~(\ref{J2}) into the first relation of (\ref{PhiTransf}) we find
\bb
\phi_\text{source}^{(1)}(x,t)=-\frac{gB_0\omega_n^2a t}{4\Omega_n^2}\sin\left(\frac{\lambda L}{2}\right)\sin\left(\Omega_n(t-x)+\frac{\lambda L}{2}\right), \label{phisource}
\ee
{ where we have assumed $t\gg x$}.

Now in the regi\'on $L\leq x\leq 2L$ we expect the ALP field to source a photon field, $A^r(x,t)$. Because of the boundary conditions, we give the following ansatz 
\bb
A^r(x,t)=\sum_{n'=1}^\infty A_{n'}'(t)\sin\omega_{n'}x
\ee
Where $\omega_n'=n' \pi/L$. We replace it into eq.~(\ref{1}), where we also consider a factor that takes into account the quality of the optical resonator in the regeneration region
\bb
\left(\partial_t^2+\gamma'\partial_t+\omega_n^2\right)A'_n(t)=\frac{(gB_0\omega_n)^2\lambda La}{16\Omega_n}\left(\frac{2}{\lambda L}\sin\left(\frac{\lambda L}{2}\right)\right)^2t\left[\sin(\omega_nt)+\sin(\Omega_n+\lambda)t\right]. \label{eqA'n}
\ee
{Note that in the last equation we have only considered the resonant mode $n=\omega L/\pi$}.
Here $\gamma'=\omega_n/Q_r$ and $Q_r$ is the quality factor in the regeneration side. Neglecting non-resonant terms we finally find the regenerated photon field in the second region, given by
\bb
A^r(x,t)=-\frac{(gB_0)^2Q_r\lambda La}{16\Omega_n}\left(\frac{2}{\lambda L}\sin\left(\frac{\lambda L}{2}\right)\right)^2t\cos\left(\omega_n t\right)\sin\left(\omega_nx\right). \label{solA'}
\ee

Now we are able to compute the probability that a photon in the first production cavity will be regenerated in the second one with the setup described. Using eq~(\ref{probability}) we find
\bb
P_{\gamma\rightarrow \phi\rightarrow \gamma}=\frac{1}{16}\left(\frac{gB_0}{\omega}\right)^4\left(\lambda L\right)^2Q_pQ_r\left|\frac{2}{\lambda L}\sin\left(\frac{\lambda L}{2}\right)\right|^4\left(\omega t\right)^2. \label{prob2}
\ee
Where we have dropped the subindex in the frequency because in resonance $\omega=\omega_n$.
As we can see, the probability shows an enhancement over time when considering a time-dependent magnetic field. Although the initial conditions in the common light-shining-through-walls-like setup are a bit different, seems instructive to compare both results. As explained in the introduction, the probability that a photon traveling into a region of extent $L$, where a static, constant magnetic field, $B_0$, is present, can be converted into an ALP and then reconverted again in a second identical region is given by
\bb
P^{static}_{\gamma\rightarrow \phi\rightarrow \gamma}=  \frac{1}{16} \frac{\omega^2}{k_{\phi}^2} \left(g B_0 L\right)^4 \frac{\mathcal F_r}{\pi} \frac{\mathcal F_p}{\pi} \left| \frac{2}{qL}\sin\left(\frac{qL}{2} \right)\right|^4.
\ee
In the limit $m_{\phi}\ll \omega$, and recalling that in resonance $\lambda\approx m_{\phi}^2/(2\omega)=q$, we see that potential gain in probability would be of 
\bb
\frac{P^{res}_{\gamma\rightarrow \phi\rightarrow \gamma}} {P^{static}_{\gamma\rightarrow \phi\rightarrow \gamma}}=\left({q t}\right)^2=7.4 \times 10^{12} \left[ \left(\frac{m_{\phi}}{10^{-6}\,\mbox{eV}}\right)^2\cdot \left( \frac{t}{3600\,\mbox{s}}\right) \right]^2. \label{estimate}
\ee
This enhancement  translates into a  gain in sensitivity of  three orders of magnitude for the coupling constant $g$, keeping in mind that a measuring time of 1 hour is a conservative estimate, according to the upgrades envisaged for these kind of experiments \cite{arias}.

\section{Discussion and outlook}\label{conclusions}
Now it seems timely to discuss the range of validity of our result, because it could seem that the probability of conversion between photons and ALPs violates unitarity. In order to find expression (\ref{prob2}) we have made the following assumptions, that we now discuss:

\begin{itemize}
\item[ i)] The frequency of the external magnetic field (tuned to the momentum transfer) should satisfy $1\ll \lambda t $, where $t$ is the measuring time. By looking at eq.~(\ref{estimate}) we see that in the mass range that can be interesting to probe with such techniques ($10^{-8}-10^{-4})$~eV, the approximation seems to hold safely. 
\item[ii)] In our perturbative approach we have only considered the quadratic term in the expression for the ALP potential (\ref{potential}). It is expected that including the non-linearity in the equations of motion the probability would appear correctly bounded. However, for the purposes of the present work, the approximated result found in eq.~(\ref{prob2}) seems to suffice. Therefore, our result is valid as soon as $\phi/f_\phi \ll1$ is satisfied.
\end{itemize}
Therefore, by the arguments raised above, we have that the two conditions that should be mutually satisfied are $\lambda t\gg1$ and $\phi/f_{\phi}\ll 1$. The second condition, by taking the amplitude of the ALP in the regeneration side can be cast as
\eqb
\frac{\phi}{f_\phi} &\approx& \frac{g^2   B_0 \pi t}{2\alpha}|a|\\
&\sim& 10^{-12} \left(\frac{g}{10^{-10}\,\mbox{GeV}^{-1}}\right)^2  \left(\frac{B_0}{10\,\mbox{T}}\cdot \frac{t}{100\,\mbox{h}}\cdot \frac{1\,\mbox{eV}}{\omega}\right) \left[ \left(\frac{\mathcal P_{p}}{1\,\mbox{W}}\right)\left(\frac{\mathcal F_p}{10^4}\right)\left(\frac{10^{-7}\,\mbox{m}^2}{S}\right) \right]^{1/2}.  \notag\eqf
Where we have used eq.~(\ref{a}). Therefore our assumptions are safe.

As a last note on the validity of our result, let us comment about neglecting the electric field sourced by currents in the walls of the cavity (conductor) in our analysis.  This electric field could potentially interfere with our resonant mode, if enhanced by the cavity. Therefore, seems necessary to impose the condition $\lambda \lesssim \pi/L$, which sets a further constraint on the mass range that could be probed with the presented setup: $m_\phi\lesssim10^{-6}\text{eV}\sqrt{n} \left[\frac{1\,\rm{m}}{L}\right].$

To summarise, we have explored the effects of a time-dependent magnetic field on the conversion between photons and new light pseudo-scalars such as axions and axion-like particles. We have found that under appropriate initial conditions of the photon field inside the cavity, it is possible to find an enhancement on the regeneration probability, that grows linearly on time. The resonant condition for this enhancement is to tune the external magnetic field to the ALP momentum transfer $\lambda\sim m_{\phi}^2/(2\omega)$. For ALP masses in scope of the cavity experiments, this frequency is of the order of GHz.  Let us note that the enhancement found here can be added to  existent enhancement methods, such as the use of  resonant cavities \cite{sikivie-bibber}.

It seems worth to explore further the potential of time-dependent electromagnetic fields to enhance the conversion probability between photons and ALPs, such as the study of the full three-dimensional problem and take into account experimental details of the setup. Although, it is beyond the scope of this work to foresee the feasibility of such modification in an experimental design.

We would like to thank the anonymous referee for useful comments. P. A. acknowledges the hospitality of Deutsches Elektronen-Synchrotron (DESY). This work was supported by FONDECYT/Chile grants 1161150 (P. A.), 1130020 (J. G.), and  Conicyt-21120890 and USA-1555 (A. A.).

\end{document}